\documentclass[12pt]{article}
\usepackage{amsmath}
\usepackage{amsfonts}
\usepackage{graphicx}
\usepackage{enumerate}
\usepackage{natbib}
\usepackage{url} 
\usepackage{threeparttable}
\usepackage{booktabs}
\usepackage{rotating}
\usepackage{afterpage}

\newtheorem{theorem}{Theorem}
\newtheorem{example}{Example}
\newtheorem{assumption}{Assumption}
\newtheorem{remark}{Remark}
\newtheorem{corollary}{Corollary}
\newtheorem{lemma}{Lemma}
\DeclareMathOperator{\IC}{IC}
\DeclareMathOperator{\diag}{diag}
\DeclareMathOperator{\interior}{int}

\begin{document}

\def\spacingset#1{\renewcommand{\baselinestretch}%
{#1}\small\normalsize} \spacingset{1}

  \title{\sc \Large PENALIZED QUASI-LIKELIHOOD ESTIMATION\\
AND MODEL SELECTION IN TIME SERIES\\MODELS
WITH PARAMETERS\\ON THE BOUNDARY\bigskip}
  \author{{\sc Heino Bohn Nielsen and Anders Rahbek}\thanks{Contact: {\tt heino.bohn.nielsen@econ.ku.dk} or {\tt anders.rahbek@econ.ku.dk}. This research was supported by the Danish Council for Independent Research (DSF Grant 7015-00028).}\bigskip\\Department of Economics, University of Copenhagen, Denmark
    }
  \maketitle

\bigskip
\begin{abstract}
\noindent We extend the theory from Fan and Li\ (2001) on penalized
likelihood-based estimation and model-selection to statistical and
econometric models which allow for non-negativity constraints on some 
or all of the parameters, as well as time-series dependence. 
It differs from classic non-penalized likelihood estimation, where limiting distributions of likelihood-based estimators and test-statistics are non-standard, and depend on the unknown number of parameters on the boundary of the parameter space. 
Specifically, we establish that the
joint model selection and estimation, results in standard asymptotic
Gaussian distributed estimators. The results are applied to the rich class
of autoregressive conditional heteroskedastic (ARCH) models for the
modelling of time-varying volatility. We find from simulations that the
penalized estimation and model-selection works surprisingly well even for a
large number of parameters. A simple empirical illustration for stock-market
returns data confirms the ability of the penalized estimation to select ARCH
models which fit nicely the autocorrelation function, as well as confirms
the stylized fact of long-memory in financial time series data.

\bigskip
\noindent%
\vfill
\end{abstract}

\newpage
\spacingset{1.2} 

\section{Introduction}

In this paper we consider penalized likelihood-based estimation of
statistical, or econometric, models parametrized by a parameter vector 
$\theta \in \mathbb{R}^{d}$, where $d$ is potentially very large and, at the
same time, where some, or all, of the entries in $\theta $ are restricted to
be non-negative. Our key interest lies in identifying the possible zero
entries in $\theta $, as well as providing an asymptotic theory for
estimation of $\theta $. To do so, we modify the theory for penalized
maximum likelihood (pML) estimation originally proposed by Fan and Li
(2001). Fan and Li (2001) consider models for independently and identically
distributed (i.i.d) data with unrestricted parameter $\theta $, in the sense
that all parameter entries are assumed to be in the interior of the
parameter space. In terms of parameters, we allow here for non-negativity
constraints, often referred to as \textquotedblleft estimation with
parameters on the boundary of the parameter space\textquotedblright , see
e.g. Andrews (1999). At the same time we extend the theory to allow for
dependence structures such as in the analysis of time series data with
time-varying mean and volatility. We provide a full asymptotic theory for
consistent model selection, in the sense that the correct zero entries in $%
\theta $ are identified as the number of observations, $n$, tends to
infinity, $n\rightarrow \infty $. Moreover, we establish conditions under
which the penalized estimator $\hat{\theta}$ is consistent, and has an
asymptotic distribution which is identical to the distribution of the
estimator if the true model was known (the so-called \textquotedblleft
oracle property\textquotedblright ). In the analysis of financial data often
quasi-likelihood estimation (QMLE)\ is applied, where the
underlying specification of the likelihood function is allowed to deviate
from the true data generating density. Our results are likewise stated for
penalized quasi-maximum likelihood (pQML).

In general, penalization is chosen to enforce sparsity, i.e. to set small
coefficients to zero, and allows simultaneous estimation and model
selection. Interestingly, the penalization of the (quasi)-likelihood
function implies that the non-zero entries of $\hat{\theta}$ are $\sqrt{n}$%
-consistent, as well as having an asymptotic Gaussian distribution. This
differs from non-penalized QMLE results, where due the non-negativity
constraints, the asymptotic limit-theory is non-standard and depends on the
number of possible parameters at the boundary, see e.g. Andrews (1999,
2001), Francq and Zako\"{\i}an (2007, 2009), Kopylev and Sinha (2010, 2011)
and Pedersen and Rahbek\ (2019).

Our results allow model-selection, or identification of the correct zero
entries of $\theta $, for even large dimensional $\theta $, which as
well-known is an inherently difficult and challenging task. Existing
approaches, which allow for non-negativity constraints, include
bootstrap-based inference, see e.g. Cavaliere, Nielsen and Rahbek (2017),
Cavaliere, Nielsen, Pedersen and Rahbek (2022) and Cavaliere, Perera and
Rahbek (2022). The bootstrap approach is based on sequential, or repeated,
testing, implying that for large dimensional $\theta $ it is computationally
highly demanding. In particular so as each single bootstrap-based test in
the sequential approach requires $2(b+1)$ nonlinear optimizations of the
likelihood, with $b$ the number of bootstrap replications.

A key example considered here is estimation and model-selection in 
time-varying volatility models applied in the analysis of financial
time series data. For volatility models it is of interest to identify the
correct volatility specification, typically with parameters subject to
non-negativity constraints. Specifically so in the vast class of
autoregressive conditional heteroskedastic (ARCH) models, and we illustrate
our theoretical results by considering simulations of different scenarios
with large dimensional parameters in ARCH models. Our focus is on
model-selection in the simulations, and we find that the discussed penalized
likelihood-based analysis works well for models with non-negativity
constraints on the parameters. Moreover, we consider an empirical
illustration in terms of daily log-returns for the Standard \& Poor's 500
index over the period 2003-2022. We investigate selection of the memory, or
lag, structure for an ARCH model with up to $d=132$ parameters, and find
that only a small fraction (1/7) are non-zero. Moreover, the results confirm the
well-known phenomenon of \textquotedblleft long-memory\textquotedblright\ in
the sense that a few significant ARCH loadings are needed at longer lags of
the conditional volatility. The resulting autocorrelation function (ACF)\
matches the empirical ACF for squared returns quite well, and is slowly
decaying, although faster than for the much applied generalized ARCH
(GARCH)\ model in Bollerslev (1986), used here as a benchmark.

The paper is organized as follows: In Section \ref{sec: Setting} we
introduce the general model set-up, and in Section \ref{sec: theory results}
we provide asymptotic theory. Section \ref{sec: sim and emp} contains
simulations as well as an empirical illustration of model selection in ARCH
models. All proofs are contained in Appendix \ref{s app}.

\section{Setting \label{sec: Setting}}

We consider here a general statistical model for the variables $\left\{
x_{t}\right\} _{t=1}^{n}$ as given by the (quasi-) log-likelihood function
denoted by $L_{n}\left( \theta \right) $, with parameter $\theta \in \Theta
\subset \mathbb{R}^{d}$. Partition $\theta $ as 
\begin{equation}
\theta =\left( \gamma ^{\prime },\beta ^{\prime }\right) ^{\prime },
\end{equation}%
where $\gamma $ and $\beta $ are of dimension $d_{\gamma },$ and $d_{\beta }$%
, respectively, with $d_{\gamma }+d_{\beta }=d$, and we let the true
parameter value be $\theta _{0}=(\gamma _{0}^{\prime },\beta _{0}^{\prime
})^{\prime }$. Here, the $d_{\beta }$ entries in $\beta $ are restricted to
be non-negative, $\beta \geq 0$, and the parameter space is thus given by,%
\begin{equation}
\Theta =\Theta _{\gamma }\times \Theta _{\beta }\text{,}
\label{def: parameter space}
\end{equation}%
where $\gamma \in \Theta _{\gamma }\subset \mathbb{R}^{d_{\gamma }}$, with $%
\Theta _{\gamma }$ compact, and $\beta \in \Theta _{\beta }=[0,\beta
_{U}]^{d_{\beta }},$ for some $\beta _{U}>0$.

As detailed below, the true value $\gamma _{0}$ for the \emph{nuisance}\
parameter $\gamma $ is known a priori to be in the interior of the parameter
space, while it is unknown for $\beta _{0}$ whether all, or some, are indeed
zero, and hence may lie on the boundary of the parameter space. Our aim is
to exploit penalization to estimate parameters, while at the same time
perform model-selection in the sense of detecting the correct number of zero
entries in $\beta $.

As a reference, the non-penalized (quasi-) maximum likelihood estimator
(QMLE)$\ \bar{\theta}$ is given by 
\begin{equation}
\bar{\theta}=\arg \max_{\theta \in \Theta }L_{n}\left( \theta \right) \text{.%
}  \label{MLEstimators}
\end{equation}%
In contrast, and reflecting that it is not known a priori whether some, or
all, of the components in $\beta _{0}$ are zero, the focus is here on
maximizing the penalized criterion function $Q_{n}\left( \theta \right) $,
given by%
\begin{equation}
Q_{n}\left( \theta \right) =L_{n}\left( \theta \right) -P_{n}\left( \beta
;\lambda \right) \text{,}  \label{eq: penalized Q}
\end{equation}%
with $\lambda $ a tuning parameter for the penalization, and penalty term $%
P_{n}\left( \beta ;\lambda \right) $ for $\beta =(\beta _{1},...,\beta
_{d_{\beta }})^{\prime }$ given by, 
\begin{equation}
P_{n}(\beta ;\lambda )=n\sum\nolimits_{j=1}^{d_{\beta }}p(\beta _{j};\lambda
)\text{.}  \label{eq: penalization P}
\end{equation}%
Note that for exposition and simplicity $p(\cdot ;\cdot )$ is set to be the
same penalty function for each $\beta _{j}$, $j=1,2,...,d_{\beta }$, and $%
\lambda \geq 0$; all theory presented covers the case where each $\beta _{j}$
has a different penalty function, $p_{j}(\beta _{j};\lambda )$ say.

For a given $\lambda $, the estimator $\hat{\theta}$ obtained by maximizing
the penalized criterion function $Q_{n}(\theta )$ is denoted the penalized
(quasi-) maximum likelihood estimator (pQMLE), 
\begin{equation}
\hat{\theta}=(\hat{\gamma}^{\prime },\hat{\beta}^{\prime })^{\prime }=\arg
\max_{\theta \in \Theta }Q_{n}(\theta )\text{.}  \label{eq: pMLE}
\end{equation}%
As to the parameter space, $\Theta =\Theta _{\gamma }\times \Theta _{\beta }$%
, where $\gamma _{0}$ is in the interior of $\Theta _{\gamma }\subset 
\mathbb{R}^{d_{\gamma }}$, while $\beta _{0}\in \Theta _{\beta }=[0,\beta
_{U}]^{d_{\beta }}$. Hence, the penalized likelihood function, $Q_{n}(\theta
)$, is differentiable in $\gamma $, and differentiable from the right for $%
\beta $ such that the penalized estimator can be found using standard
optimization algorithms.

\begin{figure}
\begin{center}
\includegraphics[width=4in]{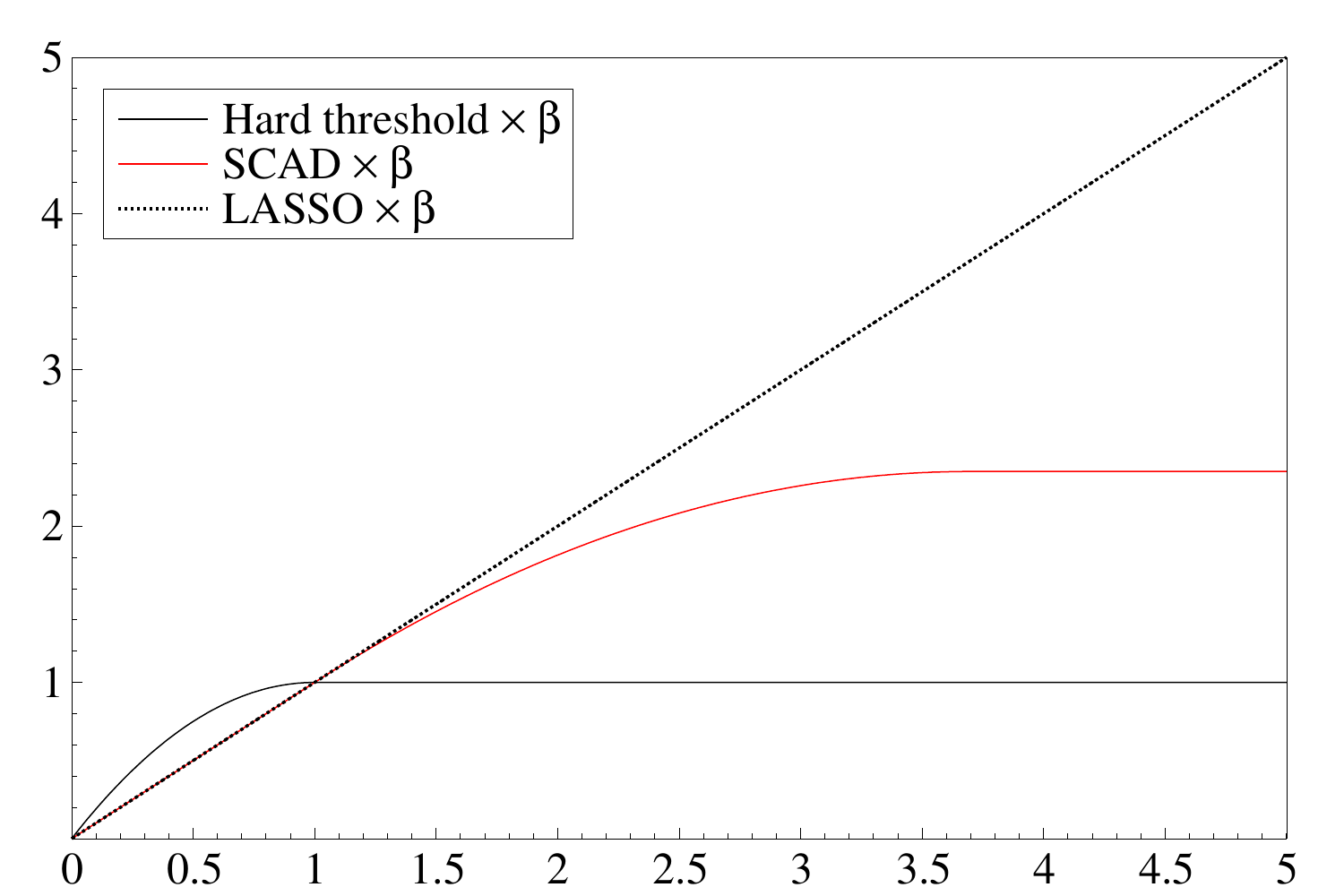}\vspace{1mm}
\end{center}
\caption{\it The LASSO, SCAD and hard threshold penalty functions with $\lambda=1$. For the SCAD penalty function the constant $a$ is set to $a=3.7$.}
\label{Fig:penalty}
\end{figure}

Specific examples of penalty functions include the classic LASSO
penalization, see Tibshirani (1996), where, as $\beta _{j}\geq 0$, the
penalty function is given by%
\begin{equation}
p(\beta _{j};\lambda )=\lambda \beta _{j}\text{.}  \label{eq: lasso}
\end{equation}%
Alternative penalty functions, as also considered in Fan and Li (2001),
include the so-called \emph{hard threshold} penalty function given by, 
\begin{equation}
p(\beta _{j};\lambda )=\lambda ^{2}-(\beta _{j}-\lambda )^{2}\mathbb{I(}%
0\leq \beta _{j}<\lambda ),  \label{eq: hardthreshold}
\end{equation}%
and the so-called \emph{smoothly clipped absolute deviation} (SCAD) penalty
function, 
\begin{equation}
p(\beta _{j};\lambda )=\lambda \beta _{j}\mathbb{I(}0\leq \beta _{j}\leq
\lambda )+\left[ \tfrac{2a\lambda \beta _{j}-\beta _{j}^{2}-\lambda ^{2}}{%
2(a-1)}\right] \mathbb{I(\lambda <}\beta _{j}\leq a\lambda )+\left[ \tfrac{%
\lambda ^{2}(a+1)}{2}\right] \mathbb{I(}\beta _{j}>a\lambda ),
\label{eq: SCAD}
\end{equation}%
where $a>2$ is a constant tuning parameter. In contrast to the LASSO in (\ref%
{eq: lasso}), the SCAD and hard threshold based pQMLE asymptotically satisfy
sparsity and the oracle-property, see Theorem \ref{thm 2} below which
extends Fan and Li (2001, Theorem 2). To illustrate Figure \ref{Fig:penalty}
plots the hard-threshold, the SCAD, and the LASSO penalty functions, from
where it follows that as $\beta _{j}$ exceeds $\lambda $, the hard threshold
and SCAD penalization vanishes, unlike for the LASSO.

\begin{example}
\label{ex: ARCH}As an example of the setting consider the ARCH($d_{\beta }$)
model given by,%
\begin{equation}
x_{t}=\mu +\varepsilon _{t}\text{, }\varepsilon _{t}=\sigma _{t}z_{t},\text{ 
}\sigma _{t}^{2}=\omega +\beta ^{\prime }v_{x,t}\text{,\quad for}\quad
t=1,....,n\text{.}  \label{eq: ARCH}
\end{equation}%
Here $\left\{ z_{t}\right\} $ is an i.i.d. sequence of standard Gaussian
variables, and the $v_{x,t}$ in the conditional variance $\sigma _{t}^{2}$
is given by 
\begin{equation*}
v_{x,t}=(\left( x_{t-1}-\mu \right) ^{2},...,\left( x_{t-d_{\beta }}-\mu
\right) ^{2})^{\prime },
\end{equation*}%
where $v_{x,1}$ is fixed in the statistical analysis. The parameter vector $%
\theta $ is here given by, $\theta =\left( \gamma ^{\prime },\beta ^{\prime
}\right) ^{\prime }$, with $\gamma =\left( \mu ,\omega \right) ^{\prime }$, $%
\mu \in \mathbb{R}$, $\omega >0$ and $\beta _{j}\geq 0$, and the Gaussian
(quasi) likelihood function is given by 
\begin{equation}
L_{n}\left( \theta \right) =-\frac{1}{2}\sum\nolimits_{t=1}^{n}\left[ \log
\sigma _{t}^{2}+\left( x_{t}-\mu \right) ^{2}/\sigma _{t}^{2}\right] \text{.}
\label{eq: ARCH lik}
\end{equation}%
For this example, it is of interest to consider estimation of $\theta $ with
the likelihood penalized for the ARCH loadings in $\beta =(\beta
_{1},..,\beta _{d_{\beta }})^{\prime }$. The penalized criterion function to
be maximized for the case of e.g. LASSO, see (\ref{eq: lasso}),\ is given by 
\begin{equation*}
Q_{n}\left( \theta \right) =L_{n}\left( \theta \right) -n\lambda
\sum\nolimits_{j=1}^{d_{\beta }}\beta _{j}\text{.}
\end{equation*}
\end{example}

\section{Theoretical Results\label{sec: theory results}}

Here we present the main results with proofs given in the appendix. To
formulate our results some notation is needed for the true value $\beta _{0}$
in order to distinguish between which components of $\beta _{0}$ are in the
interior of the parameter space, $\beta _{0,j}>0$, and, which are on the
boundary, $\beta _{0,j}=0$, $j=1,...,d_{\beta }$. We thus make the following
assumptions on the true parameter $\theta _{0}$.

\begin{assumption}[Parameter true values]
\label{Ass true parameters}With $\theta _{0}=(\gamma _{0}^{\prime },\beta
_{0}^{\prime })^{\prime }$, assume that $\gamma _{0}\in \interior\Theta
_{\gamma }$. Moreover, let $\beta _{0}^{N}=(\beta _{0,1},...,\beta
_{0,d_{N}})^{\prime }$ be the non-zero true values of $\beta _{0}$, and $%
\beta _{0}^{Z}=(\beta _{0,d_{N}+1},...\mathbin{,}\beta _{0,d_{N}+d_{Z}})^{\prime }$ be
the zero true values, with $d_{N}+d_{Z}=d_{\beta }.$
\end{assumption}

In line with Assumption \ref{Ass true parameters}, we may write $\theta
^{\prime }=\left( \gamma ^{\prime },\beta ^{\prime }\right) $, with $\beta
^{\prime }=((\beta ^{N})^{\prime },(\beta ^{Z})^{\prime })$ without loss of
generality. And we emphasize that it is alone $\beta $ which is penalized in
the likelihood analysis, while $\gamma $ is left non-penalized.

For the likelihood related quantities in terms of the process $\left\{
x_{t}\right\} _{t=1}^{n}$ we assume the following regularity conditions in
terms of (right-)derivatives of the log-likelihood function.

\begin{assumption}[Derivatives]
\label{Ass derivatives of lik}In terms of the (right-)derivatives,%
\begin{equation*}
S_{n}\left( \theta _{0}\right) =\left. \frac{\partial L_{n}\left( \theta
\right) }{\partial \theta }\right\vert _{\theta =\theta _{0}}\quad \text{and}%
\quad I_{n}\left( \theta _{0}\right) =\left. -\frac{\partial ^{2}L_{n}\left(
\theta \right) }{\partial \theta \partial \theta ^{\prime }}\right\vert
_{\theta =\theta _{0}},
\end{equation*}%
it holds, as $n\rightarrow \infty $, that:%
\begin{gather}
n^{-1/2}S_{n}\left( \theta _{0}\right) \overset{d}{\rightarrow }N\left(
0,\Omega _{S}\right) \quad \text{with\quad }\Omega _{S}>0.  \tag{R.1} \\
n^{-1}I_{n}\left( \theta _{0}\right) \overset{p}{\rightarrow }\Omega _{I}>0.
\tag{R.2} \\
\sup_{\theta \in \mathcal{N}\left( \theta _{0}\right) }\left\vert n^{-1}%
\tfrac{\partial ^{3}L_{n}\left( \theta \right) }{\partial \theta
_{i}\partial \theta _{i}\partial \theta _{k}}\right\vert \leq c_{n}\overset{p%
}{\rightarrow }c,  \tag{R.3}
\end{gather}%
for $i,j,k=1,2,...,d$, $c<\infty $ and $\mathcal{N}\left( \theta _{0}\right) 
$ some compact neighborhood of $\theta _{0}$.
\end{assumption}

Note that the regularity conditions in Assumption \ref{Ass derivatives of
lik} are identical to Assumption 1 (ii)-(iii) in Cavaliere, Nielsen,
Pedersen and Rahbek (2022), and replace conditions (A) and (B) in Fan and Li
(2001), as we extend the analysis here to allow for time series dependent
data as well as parameters on the boundary.

Observe also that, due to the shape of the parameter space, the conditions
in Assumptions 1 and 2 imply that the \emph{non-penalized} MLE $\bar{\theta}$
has a non-standard limiting distribution. Specifically, by Andrews (1999),
as $n\rightarrow \infty $,%
\begin{equation}
\sqrt{n}(\bar{\theta}-\theta _{0})\overset{d}{\rightarrow }\arg \min_{\eta
\in \mathcal{C}}q\left( \eta \right) \text{,}  \label{eq: def cone distr}
\end{equation}%
where the quadratic form $q\left( \cdot \right) $ is given by $q\left( \eta
\right) =(\eta -Z)^{\prime }\Omega _{I}(\eta -Z)$, with $Z$ distributed as $%
\Omega _{I}^{-1}N(0,\Omega _{S})$. Moreover, the quadratic form is minimized
over the cone $\mathcal{C}=\mathbb{R}^{d_{\gamma }+d_{N}}\times \mathbb{R}%
_{+}^{d_{Z}}$ where $d_{N}$ and $d_{Z}$ are defined in Assumption 1, and $%
\mathbb{R}_{+}=\mathbb{[}0,\infty )$.

\begin{remark}
\label{rem: location}To illustrate, consider the simple location model as
given by, $x_{t}=\beta +z_{t}$ with $\left\{ z_{t}\right\} _{t=1}^{n}$ an
i.i.d.(0,1) sequence and $\beta \geq 0$. With the likelihood function, $%
L_{n}\left( \beta \right) =-\frac{1}{2}\sum_{t=1}^{n}\left( x_{t}-\beta
\right) ^{2}$, the QMLE is given by $\bar{\beta}=\arg \min_{\eta \geq
0}\left( \eta -\bar{x}\right) ^{2}=\bar{x}\mathbb{I}\left( \bar{x}>0\right) $%
, where $\bar{x}=n^{-1}\sum_{t=1}^{n}x_{t}$. It holds that $\sqrt{n}(\bar{%
\beta}-\beta _{0})\overset{d}{\rightarrow }\arg \min_{\eta \in \mathcal{C}%
}\left( \eta -Z\right) ^{2}$, where the cone $\mathcal{C=}\mathbb{R}_{+}$
for $\beta _{0}=0$, while $\mathcal{C}=\mathbb{R}$ for $\beta _{0}>0$. In
contrast, the LASSO pQMLE, $\hat{\beta}_{\text{lasso}}$, and the hard
threshold pQMLE, $\hat{\beta}_{\text{hard}}$, with penalty terms in (\ref%
{eq: lasso})$\ $and (\ref{eq: hardthreshold}) respectively, are given by 
\begin{equation*}
\hat{\beta}_{\text{lasso}}=\left( \bar{x}-\lambda \right) \mathbb{I}\left( 
\bar{x}>\lambda \right) \text{ \ \ and \ \ }\hat{\beta}_{\text{hard}}=\bar{x}%
\mathbb{I}\left( \bar{x}>\lambda \right) \text{.}
\end{equation*}%
This highlights the penalty induced sparsity in the sense that the estimates
are set to zero, or at the boundary for \textquotedblleft small $\beta _{0}$%
\textquotedblright . On the other hand, the penalization for
\textquotedblleft large and positive $\beta _{0}$\textquotedblright\
vanishes asymptotically for $\hat{\beta}_{\text{hard}}$ (and for $\hat{\beta}%
_{\text{lasso}}$, provided $\lambda \rightarrow 0$). Note in this respect,
in terms of limiting theory, that the penalty parameter $\lambda $ must
satisfy $\lambda \rightarrow 0$ as $n\rightarrow \infty $, and at the same
time $\sqrt{n}\lambda \rightarrow \infty $ for both sparsity and the oracle
property to hold, see Theorems \ref{thm 1} and \ref{thm 2} below.
\end{remark}

Finally we make the following assumptions on the penalty function $p\left(
\cdot ;\cdot \right) $ from Fan and Li (2001).

\begin{assumption}[Penalty function]
\label{Ass Penalty Second order deriv}Assume that for $\lambda \rightarrow 0$
as $n\rightarrow \infty $:%
\begin{equation}
\text{(i):\ }v_{n}=\max_{j=1,2,...,d_{N}}\left\{ \left\vert \left. \tfrac{%
\partial ^{2}p(b;\lambda )}{\partial b^{2}}\right\vert _{b=\beta
_{0,j}^{N}}\right\vert \right\} \rightarrow 0\text{.}  \label{tendtozero}
\end{equation}%
\begin{equation}
\text{(ii): }\lim \inf_{n\rightarrow \infty }\lim \inf_{b\rightarrow
0}\left\{ \lambda ^{-1}\tfrac{\partial p(b;\lambda )}{\partial b}\right\} >0%
\text{.}  \label{sign}
\end{equation}
\end{assumption}

First we show (local) consistency of the penalized estimator. This is a
generalization of Fan and Li (2001, Theorem 1) to time series data with
parameters potentially on the boundary.

\begin{theorem}[Consistency]
\label{thm 1}Under Assumptions \ref{Ass true parameters}, \ref{Ass
derivatives of lik}, and \ref{Ass Penalty Second order deriv}(i), there
exists a local maximizer $\hat{\theta}$, which satisfies, 
$$\left\Vert \hat{%
\theta}-\theta _{0}\right\Vert =O_{p}(\alpha _{n}),$$
where $\alpha
_{n}=n^{-1/2}+a_{n}$ and $a_{n}=\max_{j=1,2,...,d_{N}}\left\{ \left\vert
\left. \partial p(b;\lambda )/\partial b\right\vert _{b=\beta
_{0,j}^{N}}\right\vert \right\} $.
\end{theorem}

For the hard threshold function $p\left( b;\lambda \right) =\lambda
^{2}-(b-\lambda )^{2}\mathbb{I(}b<\lambda )$ it follows that the derivatives
(from the right) with respect to $b$ are given by, 
\begin{equation*}
\frac{\partial p(b;\lambda )}{\partial b}=2(\lambda -b)\mathbb{I(}b<\lambda )%
\text{ \ \ and \ }\frac{\partial ^{2}p(b;\lambda )}{\partial b^{2}}=-2%
\mathbb{I(}b<\lambda ).
\end{equation*}%
Likewise for the SCAD penalty function, where%
\begin{equation*}
\frac{\partial p(b;\lambda )}{\partial b}=\lambda \mathbb{I(}b\leq \lambda )+%
\frac{(a\lambda -b)^{+}}{(a-1)}\mathbb{I(}b>\lambda )\text{ \ and \ \ \ }%
\frac{\partial ^{2}p(b;\lambda )}{\partial b^{2}}=\frac{1}{1-a}\mathbb{%
I(\lambda <}b\leq a\lambda )\text{.}
\end{equation*}%
Hence, for the hard threshold and SCAD penalty functions, both $a_{n}$ and $%
v_{n}$ are equal to $0$ for $\lambda \rightarrow 0$ (and $n\,$large enough).
On the other hand, for LASSO $a_{n}=\lambda $ such that $\alpha
_{n}=n^{-1/2} $ requires that $\lambda =O(n^{-1/2})$.

The next theorem states the sparsity result as well as the asymptotic
distribution, and is a generalization of Fan and Li (2001, Theorem 2).

\begin{theorem}[Asymptotic distribution]
\label{thm 2}Assume that Assumptions \ref{Ass true parameters}--\ref{Ass
Penalty Second order deriv} hold. With $\delta =(\gamma ^{\prime },\beta
^{N\prime })^{\prime }$, and $d_{\delta }=d_{\gamma }+d_{N}$, then if $%
n^{1/2}\lambda \rightarrow \infty $ and $\lambda \rightarrow 0$ as $%
n\rightarrow \infty $, with probability tending to one, $\hat{\beta}^{Z}=0$
and%
\begin{equation}
\sqrt{n}[(\hat{\delta}-\delta _{0})+(\Omega _{I,\delta }+\Sigma _{\delta
})^{-1}\mathbf{d}]\overset{d}{\rightarrow }N\left( 0,\Omega _{\delta
}\right) \text{.}  \label{eq: limit dist}
\end{equation}%
Here $\Omega _{\delta }=(\Omega _{I,\delta }+\Sigma _{\delta })^{-1}\Omega
_{S,\delta }(\Omega _{I,\delta }+\Sigma _{\delta })^{-1}$, with $\Omega
_{I,\delta }=K\Omega _{I}K^{\prime }\,$and $\Omega _{S,\delta }=K\Omega
_{S}K^{\prime }$ where $K$ is the selection matrix given by $K=\left(
I_{d_{\delta }},0_{d_{\delta }\times d_{Z}}\right) $. Moreover, the
asymptotic bias term $\mathbf{d}$ is given by, $\mathbf{d}=(0_{1\times
d_{\gamma }},\{\left. \partial p\left( b;\lambda \right) /\partial
b\right\vert _{b=\beta _{0,j}^{N}}\}_{j=1,...,d_{N}})^{\prime }$, and the
correction term, $\Sigma _{\delta }$, by $\Sigma _{\delta }=\diag%
(0_{1\times d_{\gamma }},\{\left. \partial ^{2}p\left( b;\lambda \right)
/\partial b^{2}\right\vert _{b=\beta _{0,j}^{N}}\}_{j=1,...,d_{N}})$.
\end{theorem}

Observe that for the hard threshold and the SCAD penalty function, the
result holds with $\mathbf{d}=0$ and $\Sigma _{\delta }=0$, leading to the
oracle property, i.e. that the limiting distribution is the same as for the
estimator imposing the true zero coefficients in $\beta _{0}$. For the
LASSO, however, $a_{n}=\lambda $, and asymptotic unbiasedness requires $%
\sqrt{n}\lambda \rightarrow 0$, which violates the assumptions on $\lambda $
in Theorem \ref{thm 2}.

We state the results for the hard threshold and SCAD as a corollary:

\begin{corollary}
\label{cor: 1}Consider the penalized estimation with penalty function $%
p\left( \beta _{j};\lambda \right) $ given by the hard threshold or the SCAD
in (\ref{eq: hardthreshold})\ and (\ref{eq: SCAD}). Under Assumptions \ref%
{Ass true parameters} and \ref{Ass derivatives of lik}, and with $%
n^{1/2}\lambda \rightarrow \infty $ and $\lambda \rightarrow 0$ as $%
n\rightarrow \infty $, it holds, with probability tending to one, that $\hat{%
\beta}^{Z}=0$ and%
\begin{equation*}
\sqrt{n}(\hat{\delta}-\delta _{0})\overset{d}{\rightarrow }N\left( 0,\Omega
_{I,\delta }^{-1}\Omega _{S,\delta }\Omega _{I,\delta }^{-1}\right) ,
\end{equation*}%
with $\Omega _{I,\delta }$ and $\Omega _{S,\delta }$ given in Theorem \ref%
{thm 2}.
\end{corollary}

The results in Theorem \ref{thm 2} and Corollary \ref{cor: 1} depend on the
value of $\lambda $. In practice, see also Ahrens, Hansen, and Schaffer
(2020), the choice of $\lambda $ is typically chosen by minimizing some
information criterion, $\IC(\lambda )$, with respect to $\lambda $%
. The classic Akaike (AIC), Hannan-Quinn (HQIC) and Bayesian (BIC)\
information criteria, we can state as,%
\begin{equation}
\IC(\lambda )=-2L_{n}(\hat{\theta}_{\lambda })+\hat{d}_{\delta
,\lambda }g_{n}.  \label{eq: IC}
\end{equation}%
The term $g_{n}$ takes the values $2$, $2\log \log \left( n\right) $ and $%
\log \left( n\right) $ for the AIC, HQIC\ and BIC\ respectively, and we use
the subscript $\lambda $ on $\hat{\theta}_{\lambda }$ to emphasize the
dependence of the pQMLE on $\lambda $. Moreover, $\hat{d}_{\delta ,\lambda }$
denotes the estimated number of non-zero parameters in $\hat{\theta}%
_{\lambda }$, i.e. $\hat{d}_{\delta ,\lambda }=d_{\gamma }+\dim (\hat{\beta}%
_{\lambda }^{N})$, see also Theorem \ref{thm 2}.

Specifically, for a pre-specified grid $\Lambda $ with $m$ values of $%
\lambda $, $\Lambda =\{\lambda _{1},...,\lambda _{m}\}$, $\lambda
_{i}<\lambda _{i+1}$, then $\lambda $ is chosen as $\arg \min_{\lambda \in
\Lambda }\IC\left( \lambda \right) $. Typically, $\lambda _{1}=0$
which corresponds to non-penalized QMLE, while $\lambda _{m}$ is set such
that all components of $\beta $ equal zero, $\hat{\beta}_{\lambda _{m}}=0$.

\section{Simulations and an Empirical Illustration\label{sec: sim and emp}}

To illustrate the results, we consider here model-selection in ARCH models
extended to include possible covariates in the conditional volatility,
so-called ARCH-X models, see e.g. Han and Kristensen (2014) and Pedersen and
Rahbek (2019). In terms of penalized estimation, we consider both the LASSO
and the SCAD penalizations, with focus on the sparsity result in Theorem \ref%
{thm 2}, that is, to what degree the zero entries in penalized estimators $%
\hat{\beta}$ correspond to the true zero entries of $\beta _{0}$. In
addition, in Section \ref{sec: emp} we illustrate the approach by applying
penalized model-selection and estimation to the Standard \&\ Poor's 500
index.

\subsection{Simulations Design}

The ARCH-X model is an extension of (\ref{eq: ARCH}) as given by%
\begin{equation}
x_{t}=\mu +\varepsilon _{t},\quad \varepsilon _{t}=\sigma _{t}z_{t},\quad
\sigma _{t}^{2}=\omega +\beta ^{\prime }v_{t}\text{,\quad }t=1,....,n,
\label{eq: ARCH X}
\end{equation}%
with $\left\{ z_{t}\right\} $ an i.i.d. $N\left( 0,1\right) $ sequence, and $%
v_{t}=(v_{x,t}^{\prime },v_{y,t}^{\prime })^{\prime }$ where $v_{y,t}=\left(
y_{1,t},...,y_{q,t}\right) ^{\prime }$ contains the covariates, $y_{i,t}\geq
0$, $i=1,...,q$, while, as for the previous ARCH model, $v_{x,t}=(\left(
x_{t-1}-\mu \right) ^{2},...,\left( x_{t-p}-\mu \right) ^{2})^{\prime }$.
The parameter vector is given by $\theta =(\gamma ^{\prime },\beta ^{\prime
})^{\prime }$, $\gamma =(\mu ,\omega )^{\prime }$ with $\mu \in \mathbb{R}$, 
$\gamma >0$ and $\beta _{i}\geq 0$, $i=1,...,d_{\beta }$, with $d_{\beta
}=p+q$ possibly large.

The values of $\theta =\left( \gamma ^{\prime },\beta ^{\prime }\right)
^{\prime }$ used in the simulations of the ARCH-X$\,\ $process, $\left\{
x_{t}\right\} _{t=1}^{n}$, are chosen such that the regularity conditions
for asymptotic QMLE theory of generalized ARCH-X models in Han and
Kristensen (2014) hold. This implies in particular that Assumption \ref{Ass
derivatives of lik} holds as needed for the theory to hold here. To state
the regularity conditions, rewrite initially $\beta ^{\prime }v_{t}$ in (\ref%
{eq: ARCH X}) as%
\begin{equation*}
\beta ^{\prime }v_{t}=\alpha ^{\prime }v_{x,t}+\xi ^{\prime }v_{y,t},
\end{equation*}%
such that $\beta =\left( \alpha ^{\prime },\xi ^{\prime }\right) ^{\prime }$%
, with $\alpha =(\alpha _{1},...,\alpha _{p})^{\prime }$ and $\xi =(\xi
_{1},...,\xi _{q})^{\prime }$ the ARCH loadings. With $\omega >0$, $\alpha $
is a permissible value provided the pure ARCH$\ $process with no covariates
is stationary and ergodic with $E\left( x_{t}^{6}\right) <\infty $, or $%
\sum_{i=1}^{p}\alpha _{i}^{3}<1/\kappa $ with $\kappa =E\left(
z_{t}^{6}\right) $. For $\xi $, the entries can take any non-negative value, 
$\xi _{i}\geq 0$, provided the positive covariates $y_{i,t}$ are stationary
and ergodic, and generated independently of $x_{t}$ with $E\left(
y_{i,t}^{3}\right) <\infty $. Finally, to reduce sensitivity to scaling, we
set $E(y_{it})=1$, and $\omega =1-\sum_{i=1}^{p}\alpha
_{i}-\sum_{i=1}^{q}\xi _{i},$ such that $V\left( x_{t}\right) =1$.

In terms of $v_{y,t}$, the elements are simulated as dependent over time,
mutually uncorrelated, stationary and ergodic processes with $E\left(
y_{it}\right) =1$, which is here obtained by setting $y_{it}=u_{it}^{2}$, $%
i=1,...,q$, with $u_{t}=(u_{1t},...,u_{qt})^{\prime }$ generated from a
vector autoregression,%
\begin{equation*}
u_{t}=Au_{t-1}+\Omega ^{1/2}\eta _{t},\quad t=1,2,...,n\text{.}
\end{equation*}%
Here, $u_{0}=0$, the autoregressive matrix, $A$, is defined as $A=\rho I_{q}$%
, $\rho =0.8$, $\Omega =(1-\rho ^{2})I_{q}$, and $\eta _{t}$ is an i.i.d.$%
N(0,I_{q})$ distributed sequence.

With $d_{\gamma }=2$ and $d_{N}=6$ fixed, we report results from simulations
with $d_{\beta }=p+q\in \left\{ 12,18,24,36\right\} $, and $d_{Z}\in \left\{
6,12,18,30\right\} $\emph{. }With the ARCH loadings, $\alpha =((\alpha
^{N})^{\prime },(\alpha ^{Z})^{\prime })^{\prime }$ and $\xi =((\xi
^{N})^{\prime },(\xi ^{Z})^{\prime })^{\prime }$, we set $\alpha ^{N}=\xi
^{N}=(0.15,0.15,0.10)^{\prime }$, and we report simulations with different
sample lengths $n=500,1000\,$and $2000.$

\subsection{Penalized Estimation}

With $\theta =\left( \gamma ^{\prime },\beta ^{\prime }\right) ^{\prime }$, $%
\beta =\left( \alpha ^{\prime },\xi ^{\prime }\right) ^{\prime }$, the
penalized estimation is using the Gaussian log-likelihood $L_{n}\left(
\theta \right) $ in (\ref{eq: ARCH lik}) and $\hat{\theta}=\hat{\theta}%
_{\lambda }$ is found by maximizing 
\begin{equation*}
Q_{n}\left( \theta \right) =L_{n}\left( \theta \right)
-n\sum\nolimits_{j=1}^{d_{\beta }}p\left( \beta _{j};\lambda \right) ,
\end{equation*}%
with $p\left( \cdot ;\cdot \right) $ set equal to both the LASSO penalty
function in (\ref{eq: lasso}), and the SCAD penalty in (\ref{eq: SCAD}) (with 
$a=3.7$ as suggested in Fan and Li, 2001). For the choice of $\lambda $, we
initially apply the previously mentioned classical approach minimizing the
information criteria $\IC(\lambda $) in (\ref{eq: IC})\ based on
a grid search, see e.g. Ahrens, Hansen, and Schaffer (2020). For all
reported values, we use $m=100$ non-equidistant grid points (distributed on
a log-scale) for grids $\Lambda =\left\{ \lambda _{1},...,\lambda
_{m}\right\} $, with $\lambda _{1}=0$ and $\lambda _{m}$ such that the
penalized estimators satisfy $\hat{\beta}_{\lambda _{m}}=(\hat{\alpha}%
_{\lambda _{m}}^{\prime },\hat{\xi}_{\lambda _{m}}^{\prime })^{\prime }=0$,
i.e. $\dim \hat{\beta}_{\lambda _{m}}^{N}=0$.

As an alternative we propose a computationally faster sequential strategy.
Instead of the grid search over $\Lambda =\left\{ \lambda _{1},...,\lambda
_{m}\right\} $, the proposed algorithm utilizes that for large subsets of $%
\Lambda $, a slightly modified information criteria $\IC^{\text{m}%
}\left( \lambda \right) $ is constant, thereby reducing the number of
numerical optimizations. The modified $\IC^{\text{m}}\left( \lambda
\right) $ is given by replacing the penalized QMLE$\ \hat{\theta}_{\lambda }$
in (\ref{eq: IC}) by the so-called post-estimator, $\hat{\theta}_{\lambda }^{%
\text{p}}$, that is,%
\begin{equation}
\IC^{\text{m}}(\lambda )=-2L_{n}(\hat{\theta}_{\lambda }^{%
\text{p}})+\hat{d}_{\delta ,\lambda }g_{n}.  \label{eq: IC modified}
\end{equation}%
Post-estimation is well-known from LASSO-penalized linear regression models,
see Belloni and Chernozhukov (2011), and can be defined by two steps: Given $%
\lambda $, compute the pQMLE\ $\hat{\theta}_{\lambda }=\arg \max
Q_{n}(\theta )$. Next, given $\hat{\beta}_{\lambda }^{Z}=0$ from the first
step, compute the QMLE of $\theta $ by maximizing the non-penalized
likelihood function $L_{n}\left( \theta \right) $ over $\theta \in \Theta
_{\lambda }$. Here $\Theta _{\lambda }$ is the subset of $\Theta $ with the
restriction $\beta ^{Z}=0$ corresponding to $\hat{\beta}_{\lambda }^{Z}=0$
imposed, and the post-estimator is defined by $\hat{\theta}_{\lambda }^{%
\text{p}}=\arg \max_{\theta \in \Theta _{\lambda }}L_{n}\left( \theta
\right) $.

There are two advantages from using $\IC^{\text{m}}\left( \lambda
\right) $ and post-estimation: First, the modified information criteria $%
\IC^{\text{m}}\left( \lambda \right) $ are calculated using
potentially less biased estimates of the non-zero coefficients in $\beta
^{N} $, which from the reported simulations improves model-selection.
Secondly, the likelihood function and the information criteria are alone
functions of $\lambda $ through $\hat{d}_{\delta ,\lambda }=d_{\gamma }+\dim
\beta _{\lambda }^{N}$. That is, by definition $\IC^{\text{m}%
}\left( \lambda \right) $ is piecewise constant as a function of $\lambda $,
shifting only when $\hat{d}_{\delta ,\lambda }$, i.e. the number of non-zero
parameters, changes. The shape of $\IC^{\text{m}}\left( \lambda
\right) $ as a function of $\lambda $ makes it straightforward to locate the
value of $\hat{d}_{\delta ,\lambda }$ which minimizes $\IC^{\text{m}%
}\left( \lambda \right) $, and we implement this by minimizing $\IC%
^{\text{m}}\left( \lambda \right) $ over $\lambda $ using the \emph{golden
section search} (GSS) algorithm, see e.g. Kiefer (1953), as outlined in
Appendix \ref{app: GSS}.

\subsection{Results}

Results for $d_{\beta }=p+q=12$ are reported in Table \ref{tab:pilot} for
the three different information criteria, AIC, HQIC and BIC and $n\in
\{500,1000,2000\}$. The column labelled as \textquotedblleft False $\alpha
=0 $\textquotedblright\ reports the percentage of times when $\alpha ^{N}$
were incorrectly set to zero (FZ$_{\alpha }$), while \textquotedblleft True $%
\alpha =0$\textquotedblright\ reports the percentage when $\alpha ^{Z}$ were
correctly set to zero (CZ$_{\alpha }$). Likewise for the $\xi $-columns (FZ$%
_{\xi }$ and CZ$_{\xi }$). The column labelled \textquotedblleft Average
error\textquotedblright\ reports the \ total frequency of
misclassifications, i.e.%
\begin{equation}
\frac{\dim \left( \alpha ^{N}\right) \text{FZ}_{\alpha }+\dim \left( \alpha
^{Z}\right) (100-\text{CZ}_{\alpha })+\dim \left( \xi ^{N}\right) \text{FZ}%
_{\xi }+\dim \left( \xi ^{Z}\right) (100-\text{CZ}_{\xi })}{\dim \left(
\alpha \right) +\dim \left( \xi \right) }. \label{eq: miss}
\end{equation}%
In the rows, \textquotedblleft ARCH\ QMLE\textquotedblright\
(\textquotedblleft Oracle QMLE\textquotedblright ) stands for non-penalized
QMLE (with the correct zeroes imposed). \textquotedblleft
LASSO\textquotedblright\ and \textquotedblleft SCAD\textquotedblright\ stand
for pQMLE based on LASSO and SCAD penalty, respectively, with the grid $%
\Lambda $, $\lambda \in \Lambda $, and models selected by $\IC%
\left( \lambda \right) $. \textquotedblleft P-LASSO\textquotedblright\ and
\textquotedblleft P-SCAD\textquotedblright\ use the modified information
criterion, $\IC^{\text{m}}\left( \lambda \right) $, over the grid,
while \textquotedblleft PGSS-LASSO\textquotedblright\ and \textquotedblleft
PGSS-SCAD\textquotedblright\ are based on $\IC^{\text{m}}\left(
\lambda \right) $ and GSS. Finally, we report results for the
computationally intensive -- exhaustive -- procedure, where model selection
is performed by classic AIC, HQIC$\ $and BIC,\ respectively, over all $%
2^{d_{\beta }}=4096$ candidate models. This we refer to as an exhaustive
search.

First, the simulation results illustrate that model selection based on the
SCAD\ penalization is superior to results for the LASSO, which is in line
with the superior theoretical results for the SCAD in terms of scarcity and
oracle properties.

Next, we find the general result that LASSO and SCAD performs less well when
compared to P-LASSO and P-SCAD, indicating that post-estimation improves
model-selection. Interestingly, post-estimation based model selection for
LASSO and SCAD are comparable. In addition, result for the PGSS-LASSO
(PGSS-SCAD) are very close to the results for the grid search. At the same
time, the results based on the exhaustive search and PGSS-LASSO\ (PGSS-SCAD)
are very close, indicating that the much less computationally demanding GSS
versions are to be preferred. In particular so as the exhaustive search in
larger models (i.e. with $d_{\beta }$ even moderately large) are
computationally infeasible. As an example, Table \ref{tab:zerosVARB} reports
results for $d_{\beta }\in \{18,24,36\}$ corresponding to $262144$, $%
16777216 $ and $68719476736$ numerical optimizations for the -- in these
cases -- infeasible exhaustive search\emph{.}

Based on the average error, there is a tendency for model selection based on
the AIC to dominate in the small sample ($n=500$), while HQIC and BIC
dominate in medium and large samples, ($n=1000$ or $n=2000$). Finally,
observe that in large samples, model selection based on the BIC criterion is
close to the results for the infeasible oracle estimator.

Table \ref{tab:zerosVARB} reports results for the BIC information criterion
(results for the AIC\ and HQIC were similar) for $d_{\beta }\in \left\{
18,24,36\right\} $, and we find again that penalized estimation, in
combination with post-estimation and GSS,\ works remarkably well. In
particular, we note that the results are highly robust to the value of $%
d_{\beta }$. Finally, unreported results show the same fine findings when
introducing non-diagonal $\Omega $ and hence contemporaneously correlated $%
v_{y,t}$.

{
\renewcommand{\baselinestretch}{1.0}%
\begin{sidewaystable}[p] \par\centering \footnotesize\addvspace{-0mm}
\hspace*{-6mm}
\begin{threeparttable}[b]

\caption{Model selection based on three infomation critiria with $d_{N} = d_{Z}=6$ and $n\in\{500,1000,2000\}$.}

\begin{tabular}{lllrrrrrlrrrrrlrrrrr}
\toprule
&  &    & \multicolumn{5}{c}{$n=500$} &    & \multicolumn{5}{c}{$n=1000$} &    & \multicolumn{5}{c}{$n=2000$} \\
\cmidrule{4-8}\cmidrule{10-14}\cmidrule{16-20}
Estimator& IC   &    & \multicolumn{1}{c}{False} & \multicolumn{1}{c}{True} & \multicolumn{1}{c}{False} & \multicolumn{1}{c}{True} & \multicolumn{1}{c}{Average} &    & \multicolumn{1}{c}{False} & \multicolumn{1}{c}{True} & \multicolumn{1}{c}{False} & \multicolumn{1}{c}{True} & \multicolumn{1}{c}{Average} &    & \multicolumn{1}{c}{False} & \multicolumn{1}{c}{True} & \multicolumn{1}{c}{False} & \multicolumn{1}{c}{True} & \multicolumn{1}{c}{Average} \\
&    &    & \multicolumn{1}{c}{$\alpha=0$} & \multicolumn{1}{c}{$\alpha=0$} & \multicolumn{1}{c}{$\xi=0$} & \multicolumn{1}{c}{$\xi=0$} & \multicolumn{1}{c}{error} &    & \multicolumn{1}{c}{$\alpha=0$} & \multicolumn{1}{c}{$\alpha=0$} & \multicolumn{1}{c}{$\xi=0$} & \multicolumn{1}{c}{$\xi=0$} & \multicolumn{1}{c}{error} &    & \multicolumn{1}{c}{$\alpha=0$} & \multicolumn{1}{c}{$\alpha=0$} & \multicolumn{1}{c}{$\xi=0$} & \multicolumn{1}{c}{$\xi=0$} & \multicolumn{1}{c}{error} \\
\cmidrule{1-2}\cmidrule{4-8}\cmidrule{10-14}\cmidrule{16-20}    ARCH QMLE &    &    & 1.9 & 61.5 & 0.8 & 52.6 & 22.1 &    & 0.2 & 59.1 & 0.0 & 52.1 & 22.3 &    & 0.0 & 58.4 & 0.0 & 50.1 & 22.9 \\
Oracle QMLE &    &    & 1.4 & 100.0 & 0.7 & 100.0 & 0.5 &    & 0.2 & 100.0 & 0.0 & 100.0 & 0.1 &    & 0.0 & 100.0 & 0.0 & 100.0 & 0.0 \\
\cmidrule{1-2}\cmidrule{4-8}\cmidrule{10-14}\cmidrule{16-20}
&    &    &    &    &    &    &    &    &    &    &    &    &    &    &    &    &    &    &  \\
LASSO & AIC &    & 3.5 & 71.8 & 2.2 & 69.6 & 16.0 &    & 0.3 & 68.3 & 0.1 & 67.8 & 16.1 &    & 0.0 & 67.0 & 0.0 & 66.4 & 16.7 \\
SCAD & AIC &    & 7.2 & 79.9 & 3.5 & 76.1 & 13.7 &    & 1.1 & 79.8 & 0.4 & 78.1 & 10.9 &    & 0.0 & 83.2 & 0.0 & 81.3 & 8.9 \\
&    &    &    &    &    &    &    &    &    &    &    &    &    &    &    &    &    &    &  \\
P-LASSO & AIC &    & 10.8 & 89.6 & 9.5 & 90.2 & 10.1 &    & 2.1 & 90.9 & 1.6 & 92.8 & 5.0 &    & 0.0 & 91.7 & 0.1 & 92.5 & 4.0 \\
P-SCAD & AIC &    & 11.9 & 90.4 & 9.1 & 89.8 & 10.2 &    & 2.4 & 91.8 & 1.5 & 92.5 & 4.9 &    & 0.0 & 91.9 & 0.0 & 92.2 & 4.0 \\
&    &    &    &    &    &    &    &    &    &    &    &    &    &    &    &    &    &    &  \\
PGSS-LASSO & AIC &    & 10.2 & 88.0 & 9.1 & 89.5 & 10.4 &    & 1.6 & 89.6 & 1.6 & 92.3 & 5.3 &    & 0.0 & 91.0 & 0.1 & 92.4 & 4.2 \\
PGSS-SCAD & AIC &    & 11.2 & 89.3 & 8.8 & 89.0 & 10.4 &    & 2.0 & 90.6 & 1.6 & 91.9 & 5.3 &    & 0.0 & 91.3 & 0.1 & 92.2 & 4.2 \\
&    &    &    &    &    &    &    &    &    &    &    &    &    &    &    &    &    &    &  \\
Exhaustive & AIC &    & 16.0 & 93.2 & 8.8 & 90.8 & 10.2 &  & 2.7 & 94.0 & 1.3 & 92.4 & 4.4 &  & 0.0 & 92.8 & 0.0 & 91.9 & 3.8 \\
\cmidrule{1-2}\cmidrule{4-8}\cmidrule{10-14}\cmidrule{16-20}
&    &    &    &    &    &    &    &    &    &    &    &    &    &    &    &    &    &    &  \\
LASSO & HQIC &    & 5.7 & 76.3 & 4.4 & 76.5 & 14.3 &    & 0.4 & 72.5 & 0.3 & 74.2 & 13.5 &    & 0.0 & 71.4 & 0.0 & 74.3 & 13.6 \\
SCAD & HQIC &    & 9.0 & 82.9 & 5.3 & 81.0 & 12.6 &    & 1.5 & 82.6 & 0.7 & 83.5 & 9.0 &    & 0.1 & 85.8 & 0.1 & 86.9 & 6.9 \\
&    &    &    &    &    &    &    &    &    &    &    &    &    &    &    &    &    &    &  \\
P-LASSO & HQIC &    & 16.6 & 92.3 & 17.4 & 94.7 & 11.8 &    & 4.6 & 94.9 & 4.1 & 97.8 & 4.0 &    & 0.2 & 96.8 & 0.2 & 98.0 & 1.4 \\
P-SCAD & HQIC &    & 19.1 & 94.0 & 17.6 & 95.3 & 11.9 &    & 4.8 & 95.2 & 3.9 & 97.5 & 4.0 &    & 0.2 & 97.0 & 0.2 & 97.9 & 1.4 \\
&    &    &    &    &    &    &    &    &    &    &    &    &    &    &    &    &    &    &  \\
PGSS-LASSO & HQIC &    & 18.4 & 93.7 & 18.2 & 95.5 & 11.9 &    & 3.7 & 93.2 & 4.1 & 97.4 & 4.3 &    & 0.2 & 95.3 & 0.2 & 97.8 & 1.8 \\
PGSS-SCAD & HQIC &    & 17.2 & 92.8 & 17.2 & 94.4 & 11.8 &    & 4.0 & 93.8 & 4.0 & 97.2 & 4.3 &    & 0.2 & 95.6 & 0.2 & 97.7 & 1.8 \\
&    &    &    &    &    &    &    &    &    &    &    &    &    &    &    &    &    &    &  \\
Exhaustive & HQIC &    & 18.4 & 94.6 & 10.6 & 92.4 & 10.5 &  & 3.2 & 95.6 & 1.5 & 94.4 & 3.7 &  & 0.1 & 94.6 & 0.1 & 93.6 & 3.0 \\
\cmidrule{1-2}\cmidrule{4-8}\cmidrule{10-14}\cmidrule{16-20}
&    &    &    &    &    &    &    &    &    &    &    &    &    &    &    &    &    &    &  \\
LASSO & BIC &    & 13.1 & 83.6 & 14.0 & 86.4 & 14.3 &    & 0.9 & 78.2 & 1.1 & 83.1 & 10.2 &    & 0.0 & 77.1 & 0.0 & 82.2 & 10.2 \\
SCAD & BIC &    & 15.2 & 86.8 & 13.2 & 87.7 & 13.5 &    & 2.0 & 85.2 & 1.6 & 88.2 & 7.6 &    & 0.1 & 88.1 & 0.1 & 90.7 & 5.4 \\
&    &    &    &    &    &    &    &    &    &    &    &    &    &    &    &    &    &    &  \\
P-LASSO & BIC &    & 28.2 & 96.9 & 33.5 & 98.3 & 16.6 &    & 8.4 & 97.0 & 9.8 & 99.5 & 5.4 &    & 1.1 & 98.2 & 0.7 & 99.6 & 1.0 \\
P-SCAD & BIC &    & 28.5 & 97.1 & 32.7 & 98.1 & 16.5 &    & 8.3 & 97.2 & 9.5 & 99.4 & 5.3 &    & 1.1 & 98.5 & 0.7 & 99.6 & 0.9 \\
&    &    &    &    &    &    &    &    &    &    &    &    &    &    &    &    &    &    &  \\
PGSS-LASSO & BIC &    & 26.6 & 96.2 & 33.5 & 97.8 & 16.5 &    & 6.8 & 95.7 & 9.2 & 99.1 & 5.3 &    & 0.7 & 96.8 & 0.8 & 99.5 & 1.3 \\
PGSS-SCAD & BIC &    & 26.9 & 96.3 & 32.9 & 97.6 & 16.5 &    & 6.8 & 95.8 & 8.9 & 99.0 & 5.2 &    & 0.7 & 96.9 & 0.8 & 99.5 & 1.3 \\
&    &    &    &    &    &    &    &    &    &    &    &    &    &    &    &    &    &    &  \\
Exhaustive & BIC &    & 21.9 & 95.7 & 12.8 & 94.6 & 11.1 &  & 4.4 & 96.8 & 2.4 & 96.5 & 3.4 &  & 0.1 & 96.8 & 0.1 & 95.9 & 1.9 \\
\bottomrule
\end{tabular}\label{tab:pilot}

\linespread{.9}\selectfont\textsc{Note:} {\it
Based on 1000 Monte Carlo replications.
The average error is calculated as in equation (\ref{eq: miss}).
}
\end{threeparttable}
\end{sidewaystable}
}

{
\spacingset{1.0}
\begin{table}[p] \par\addvspace{-15mm}\centering \footnotesize
\hspace*{-0mm}
\begin{threeparttable}[b]
\addvspace{-12mm}

\caption{Model selection based on the BIC information criterion, $d_{N} = 6$, $d_{Z} \in \{6,12,18,30\}$ for $n=1000$.}
\begin{tabular}{lrlrrrrr}
\toprule
Estimator & \multicolumn{1}{l}{$d_{\beta}$}   &    & \multicolumn{1}{c}{False} & \multicolumn{1}{c}{True} & \multicolumn{1}{c}{False} & \multicolumn{1}{c}{True} & \multicolumn{1}{c}{Average} \\
&    &    & \multicolumn{1}{c}{$\alpha=0$} & \multicolumn{1}{c}{$\alpha=0$} & \multicolumn{1}{c}{$\xi=0$} & \multicolumn{1}{c}{$\xi=0$} & \multicolumn{1}{c}{error} \\
\cmidrule{1-2}\cmidrule{4-8} 
   &    &    &    &    &    &    &  \\
ARCH QMLE & 12  &    & 0.2 & 59.1 & 0.0 & 52.1 & 22.3 \\
Oracle QMLE & 12  &    & 0.2 & 100.0 & 0.0 & 100.0 & 0.1 \\
   &    &    &    &    &    &    &  \\
LASSO & 12  &    & 0.9 & 78.2 & 1.1 & 83.1 & 10.2 \\
SCAD & 12  &    & 2.0 & 85.2 & 1.6 & 88.2 & 7.6 \\
   &    &    &    &    &    &    &  \\
P-LASSO & 12  &    & 8.4 & 97.0 & 9.8 & 99.5 & 5.4 \\
P-SCAD & 12  &    & 8.3 & 97.2 & 9.5 & 99.4 & 5.3 \\
   &    &    &    &    &    &    &  \\
PGSS-LASSO & 12  &    & 6.8 & 95.7 & 9.2 & 99.1 & 5.3 \\
PGSS-SCAD & 12  &    & 6.8 & 95.8 & 8.9 & 99.0 & 5.2 \\
\cmidrule{1-2}\cmidrule{4-8}       &    &    &    &    &    &    &  \\
ARCH QMLE & 18 &    & 0.2 & 63.1 & 0.0 & 51.3 & 28.6 \\
Oracle QMLE & 18 &    & 0.2 & 100.0 & 0.0 & 100.0 & 0.0 \\
   &    &    &    &    &    &    &  \\
LASSO & 18 &    & 1.4 & 84.8 & 1.5 & 86.7 & 10.0 \\
SCAD & 18 &    & 2.0 & 87.9 & 1.6 & 88.9 & 8.3 \\
   &    &    &    &    &    &    &  \\
P-LASSO & 18 &    & 8.5 & 97.9 & 9.0 & 99.4 & 3.8 \\
P-SCAD & 18 &    & 8.6 & 98.0 & 8.9 & 99.4 & 3.8 \\
   &    &    &    &    &    &    &  \\
PGSS-LASSO & 18 &    & 7.3 & 97.3 & 9.0 & 98.9 & 4.0 \\
PGSS-SCAD & 18 &    & 7.4 & 97.4 & 8.9 & 98.8 & 4.0 \\
\cmidrule{1-2}\cmidrule{4-8}       &    &    &    &    &    &    &  \\
ARCH QMLE & 24 &    & 0.2 & 65.8 & 0.1 & 51.6 & 31.0 \\
Oracle QMLE & 24 &    & 0.1 & 100.0 & 0.1 & 100.0 & 0.0 \\
   &    &    &    &    &    &    &  \\
LASSO & 24 &    & 1.7 & 87.8 & 2.1 & 89.4 & 9.0 \\
SCAD & 24 &    & 2.8 & 89.3 & 2.1 & 90.5 & 8.2 \\
   &    &    &    &    &    &    &  \\
P-LASSO & 24 &    & 8.6 & 98.6 & 10.4 & 99.2 & 3.2 \\
P-SCAD & 24 &    & 8.7 & 98.6 & 10.2 & 99.2 & 3.2 \\
   &    &    &    &    &    &    &  \\
PGSS-LASSO & 24 &    & 7.7 & 98.0 & 10.7 & 99.0 & 3.4 \\
PGSS-SCAD & 24 &    & 7.8 & 98.1 & 10.5 & 99.0 & 3.4 \\
\cmidrule{1-2}\cmidrule{4-8}       &    &    &    &    &    &    &  \\
ARCH QMLE & 36 &    & 0.3 & 68.7 & 0.0 & 55.2 & 31.8 \\
Oracle QMLE & 36 &    & 0.2 & 100.0 & 0.0 & 100.0 & 0.0 \\
   &    &    &    &    &    &    &  \\
LASSO & 36 &    & 2.2 & 91.4 & 3.6 & 93.3 & 6.9 \\
SCAD & 36 &    & 2.9 & 91.4 & 2.7 & 92.6 & 7.1 \\
   &    &    &    &    &    &    &  \\
P-LASSO & 36 &    & 8.7 & 98.7 & 11.3 & 99.4 & 2.5 \\
P-SCAD & 36 &    & 8.5 & 98.7 & 11.2 & 99.4 & 2.4 \\
   &    &    &    &    &    &    &  \\
PGSS-LASSO & 36 &    & 8.4 & 98.4 & 12.9 & 99.3 & 2.7 \\
PGSS-SCAD & 36 &    & 8.4 & 98.4 & 12.9 & 99.3 & 2.7 \\
\bottomrule
\end{tabular}\label{tab:zerosVARB}

\linespread{.9}\selectfont\textsc{Note:} {\it
Based on 1000 Monte Carlo replications.
The average error is calculated as in equation (\ref{eq: miss}).
}
\end{threeparttable}
\end{table}%
}

\subsection{Empirical Illustration\label{sec: emp}}

\begin{figure}
\begin{center}
\includegraphics[width=5in]{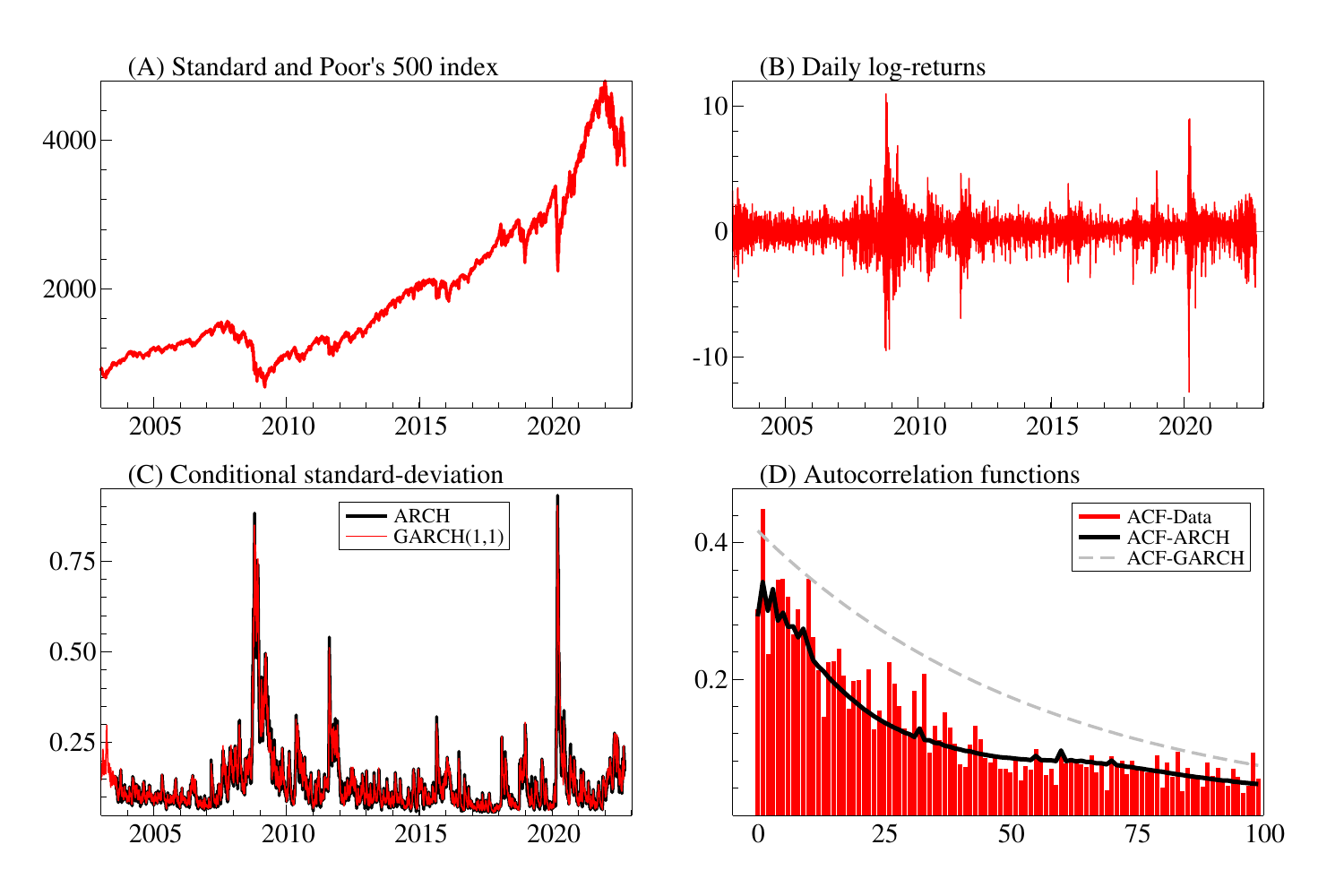}
\end{center}
\caption{\it Panel (A) plots the Standard and Poor's 500 index, and Panel (B) the log-returns, $x_t$. 
Panel (C) shows the estimated conditional volatility of the ARCH(130) model (with $d_Z=114$ zero parameters) together
with the benchmark GARCH. Panel (D) shows the empirical ACF of $x_t^2$ together with implied ARCH and GARCH ACF's respectively. 
}
\label{fig:sp500}
\end{figure}

We consider here daily observations of the Standard \& Poor's 500-index
(S\&P500) for the period 2/1-2003 to 27/9-2022, and hence $4969$
observations. In Figure \ref{fig:sp500}, Panel (A) plots the S\&P500 index
and Panel (B) the corresponding log-returns, $\left\{ x_{t}\right\}
_{t=1}^{n}$. Penalized estimation of the ARCH model in (\ref{ex: ARCH}) with 
$d_{\beta }=130$ (and hence a total of $d=132$ parameters), lead to $d_{N}=16
$ non-zero parameters, or $d_{Z}=114$ zero entries in $\sigma _{t}^{2}$, the
conditional volatility. The resulting model has significant $\beta _{i}$
entries, or ARCH loadings, for $i\in S\cup L$, with $S=\left\{
1,2,..,11\right\} $ and $L=\left\{ 33,56,61,71,120\right\} $. Here
\textquotedblleft $L$\textquotedblright\ is used for \textquotedblleft
Long\textquotedblright\ in the sense that the significant ARCH loadings $%
\left\{ \beta _{i}\right\} _{i\in L}$ indicate the stylized fact that the autocorrelation 
function for squared log-returns for financial time series show high-persistence, 
or long memory. 
Notice in this aspect from Figure \ref{fig:sp500} Panels (C) and (D), that
while the estimated conditional volatility function $\hat{\sigma}_{t}^{2}$
is very similar to that of the benchmark model in financial time series, the
generalized ARCH\ (GARCH, or GARCH(1,1)) of Bollerslev (1986), the ACF\ of
the estimated ARCH is more rapid decaying than that of the GARCH. 
We emphasize that the findings of the need for a \textquotedblleft long-memory\textquotedblright\ 
structure together 
with a more steep decay in the ACF (than that of the benchmark GARCH) do not 
change if the sample is changed by initiating the analysis at any year during 1997 to 2001.

\newpage

\appendix

\section{Proofs\label{s app}}

The proofs are based on modifying the arguments in the proofs of Fan and Li
(2001, proofs of Theorems 1 and 2) in order to allow for parameters on the
boundary of the parameter space as well as time-series dependence. In
particular, as the log-likelihood function and the penalty terms are only
differentiable from the right in the direction of $\beta $ in $\theta
=\left( \gamma ^{\prime },\beta ^{\prime }\right) ^{\prime }$, non-standard
log-likelihood expansions from Andrews (1999) are applied.

\subsection{Proof of Theorem \protect\ref{thm 1}:}

With $\mathbf{u=(u}_{\gamma }^{\prime }\mathbf{,u}_{^{N}}^{\prime },\mathbf{u%
}_{^{Z}}^{\prime })^{\prime }\in \mathbb{R}^{d_{\gamma }+d_{N}}\times 
\mathbb{R}_{+}^{d_{Z}}$, where $\mathbb{R}_{+}=[0,\infty )$, we establish,
for any $\varepsilon >0$,%
\begin{equation*}
P\left\{ \sup_{||\mathbf{u}||=c}\left[ Q_{n}\left( \theta _{0}+\alpha _{n}%
\mathbf{u}\right) <Q_{n}\left( \theta _{0}\right) \right] \right\} \geq
1-\varepsilon ,
\end{equation*}%
for $c$ large and where $\alpha _{n}=n^{-1/2}+a_{n}.$ This implies (with
probability $1-\varepsilon $) that $Q_{n}$ has a maximum $\hat{\theta}\,$in $%
\left\{ \theta :\theta =\theta _{0}+\alpha _{n}\mathbf{u,}\left\Vert \mathbf{%
u}\right\Vert \leq c\right\} \cap \Theta $, with $\left\Vert \hat{\theta}%
-\theta _{0}\right\Vert =O_{p}\left( \alpha _{n}\right) $.

By Andrews (1999, Theorem 6), the log-likelihood function has the following
expansion based on derivatives (from the right in $\beta $) at $\theta
=\left( \gamma ^{\prime },\beta ^{\prime }\right) ^{\prime }\in \Theta $, 
\begin{gather}
L_{n}\left( \theta \right) -L_{n}\left( \theta _{0}\right) =\left( \theta
-\theta _{0}\right) ^{\prime }S_{n}\left( \theta _{0}\right) -\left( \theta
-\theta _{0}\right) ^{\prime }I_{n}\left( \theta _{0}\right) \left( \theta
-\theta _{0}\right) /2+o_{p}\left( 1\right)  \label{app: expansion of lik} \\
=\left[ n^{-1}Z_{n}\left( \theta _{0}\right) I_{n}\left( \theta _{0}\right)
Z_{n}\left( \theta _{0}\right) \right] /2-q\left( n^{1/2}\left( \theta
-\theta _{0}\right) \right) /2+o_{p}\left( 1\right) .  \notag
\end{gather}%
Here the quadratic form, $q\left( x\right) =\left( x-Z_{n}\left( \theta
_{0}\right) \right) ^{\prime }\left( I_{n}\left( \theta _{0}\right)
/n\right) \left( x-Z_{n}\left( \theta _{0}\right) \right) $, with $%
Z_{n}\left( \theta _{0}\right) =\left( n^{-1}I_{n}\left( \theta _{0}\right)
\right) ^{-1}n^{-1/2}S_{n}\left( \theta _{0}\right) =O_{p}\left( 1\right) $
by Assumption \ref{Ass derivatives of lik} (\emph{R.1, R.2}), and the
remainder term in (\ref{app: expansion of lik})\ is $o_{p}\left( 1\right) $
by by Assumption \ref{Ass derivatives of lik} (\emph{R.3}). With $\theta
=\theta _{0}+\alpha _{n}\mathbf{u},$ rewrite (\ref{app: expansion of lik})
as,%
\begin{gather*}
L_{n}\left( \theta _{0}+\alpha _{n}\mathbf{u}\right) -L_{n}\left( \theta
_{0}\right) =\alpha _{n}\mathbf{u}^{\prime }S_{n}\left( \theta _{0}\right)
-\alpha _{n}^{2}\mathbf{u}^{\prime }I_{n}\left( \theta _{0}\right) \mathbf{u}%
/2+o_{p}\left( 1\right) \\
=\left[ n^{-1}Z_{n}\left( \theta _{0}\right) ^{\prime }I_{n}\left( \theta
_{0}\right) Z_{n}\left( \theta _{0}\right) \right] /2-\left( n\alpha
_{n}^{2}\right) q_{\alpha }\left( \mathbf{u}\right) /2+o_{p}\left( 1\right)
\end{gather*}%
where 
\begin{equation*}
q_{\alpha }\left( \mathbf{u}\right) =(\mathbf{u}-\left( n^{1/2}\alpha
_{n}\right) ^{-1}Z_{n}\left( \theta _{0}\right) )^{\prime }\left(
I_{n}\left( \theta _{0}\right) /n\right) (\mathbf{u}-\left( n^{1/2}\alpha
_{n}\right) ^{-1}Z_{n}\left( \theta _{0}\right) ).
\end{equation*}%
Next, by definition of the penalty function, $p\left( \beta
_{0,j}^{Z};\lambda \right) =0$, $j=d_{N}+1,...,d_{\beta }$, and hence%
\begin{align*}
Q_{n}\left( \theta _{0}+\alpha _{n}\mathbf{u}\right) -Q_{n}\left( \theta
_{0}\right) & \leq \underset{A_{n}}{\underbrace{\left[ n^{-1}Z_{n}\left(
\theta _{0}\right) I_{n}\left( \theta _{0}\right) Z_{n}\left( \theta
_{0}\right) \right] /2-\left( n\alpha _{n}^{2}\right) q_{\alpha _{n}}\left( 
\mathbf{u}\right) /2+o_{p}\left( 1\right) }} \\
& \underset{B_{n}}{\underbrace{-n\sum_{j=1}^{d_{N}}\left[ p\left( \beta
_{0,j}^{N}+\alpha _{n}\mathbf{u}_{_{N,},j};\lambda \right) -p\left( \beta
_{0,j}^{N};\lambda \right) \right] }}\text{.}
\end{align*}%
Using $Z_{n}\left( \theta _{0}\right) =O_{p}(1)$ and $I_{n}\left( \theta
_{0}\right) =O_{p}\left( n\right) $, $A_{n}=O_{p}\left( n\alpha
_{n}^{2}\right) $ and hence $\left( n\alpha _{n}^{2}\right) q_{\alpha
_{n}}\left( \mathbf{u}\right) /2$ in $A_{n}$ dominates for $\left\Vert 
\mathbf{u}\right\Vert =c$, $c$ large. Next, for the term $B_{n},$ a Taylor
expansion gives,%
\begin{equation*}
B_{n}=-n\sum\nolimits_{j=1}^{d_{N}}\left[ \alpha _{n}\partial p(\beta
_{0,j}^{N})\mathbf{u}_{\beta ^{N},j}+\alpha _{n}^{2}\partial ^{2}p(\beta
_{0,j}^{N})\mathbf{u}_{\beta ^{N},j}^{2}\left( \tfrac{1}{2}+o\left( 1\right)
\right) \right] ,
\end{equation*}%
with $\partial p(\beta _{0,j}^{N})=\left. \partial p\left( b;\lambda \right)
/\partial b\right\vert _{b=\beta _{0,j}^{N}}$ and $\partial ^{2}p\left(
\beta _{0,j}^{N}\right) =\left. \partial ^{2}p\left( b;\lambda \right)
/\partial b^{2}\right\vert _{b=\beta _{0,j}^{N}}$. Recall $%
a_{n}=\max_{j=1,2,...,d_{N}}\left\{ |\partial p(\beta _{0,j}^{N})|\right\} $%
, and hence $B_{n}$ is bounded by%
\begin{equation*}
\sqrt{d_{N}}n\alpha _{n}a_{n}\left\Vert \mathbf{u}\right\Vert +\left(
n\alpha _{n}^{2}\right) v_{n}\left\Vert \mathbf{u}\right\Vert ^{2},
\end{equation*}%
where $v_{n}\rightarrow 0$ by Assumption \ref{Ass Penalty Second order deriv}%
. Thus for $c$ sufficiently large, indeed $\left( n\alpha _{n}^{2}\right)
q_{\alpha _{n}}\left( \mathbf{u}\right) $ dominates, and the inequality
holds as desired. \hfill $\square $

\subsection{Proof of Theorem \protect\ref{thm 2}:}

By Lemma \ref{lem 1} below $\hat{\beta}^{Z}=0$, and with $\delta =(\gamma
^{\prime },(\beta ^{N})^{\prime })^{\prime }$, we get 
\begin{equation*}
\left. \partial Q_{n}\left( \theta \right) /\partial \delta \right\vert
_{\theta =\left( \hat{\delta}^{\prime },0\right) ^{\prime }}=\left. \partial
L_{n}\left( \theta \right) /\partial \delta \right\vert _{\theta =\left( 
\hat{\delta}^{\prime },0\right) ^{\prime
}}-n\sum\nolimits_{j=1}^{d_{N}}\partial p(\hat{\beta}_{j}^{N})\text{,}
\end{equation*}%
where $\partial p(\hat{\beta}_{j}^{N})=\left. \partial p\left( b;\lambda
\right) /\partial b\right\vert _{b=\beta _{0,j}^{N}}$, $j=1,...,d_{N}$. By a
Taylor expansion in terms of $\delta $%
\begin{align}
0& =\partial L_{n}\left( \theta _{0}\right) /\partial \delta -\left[
-\partial ^{2}L_{n}\left( \theta _{0}\right) /\partial \delta \partial
\delta ^{\prime }+o_{p}\left( 1\right) \right] (\hat{\delta}-\delta _{0})
\label{app: Score expansion} \\
& -n(\sum\nolimits_{j=1}^{d_{N}}\partial p\left( \beta _{0,j}^{N}\right) + 
\left[ \partial ^{2}p\left( \beta _{0,j}^{N}\right) +o_{p}\left( 1\right) %
\right] (\hat{\beta}_{j}^{N}-\beta _{0,j}^{N}))\text{.}  \notag
\end{align}%
Next, by Assumption \ref{Ass derivatives of lik}, $S_{n,\delta
}=n^{-1/2}\partial L_{n}\left( \theta _{0}\right) /\partial \delta \overset{d%
}{\rightarrow }N\left( 0,\Omega _{S,\delta }\right) $ with $\Omega
_{S,\delta }=K\Omega _{S}K^{\prime }$ where $K$ is the selection matrix
given by $K=\left( I_{d_{\delta }},0_{d_{\delta }\times d_{Z}}\right) $, and 
$I_{n,\delta }=-n^{-1}\partial ^{2}L_{n}\left( \theta _{0}\right) /\partial
\delta \partial \delta ^{\prime }\overset{p}{\rightarrow }\Omega _{I,\delta
}=K\Omega _{I}K^{\prime }$. Next, the result holds by using the identity,%
\begin{equation*}
\sqrt{n}\sum\nolimits_{j=1}^{d_{N}}\left[ \partial ^{2}p\left( \beta
_{0,j}^{N}\right) +o_{p}\left( 1\right) \right] (\hat{\beta}_{j}^{N}-\beta
_{0,j}^{N})=[\Sigma _{\delta }+o_{p}\left( 1\right) ]\sqrt{n}(\hat{\delta}%
-\delta _{0}),
\end{equation*}%
where $\Sigma _{\delta }={\diag}(0_{1\times d_{\gamma }},\{\left.
\partial ^{2}p\left( b;\lambda \right) /\partial b^{2}\right\vert _{b=\beta
_{0,j}^{N}}\}_{j=1,...,d_{N}})$, to rewrite (\ref{app: Score expansion}) as,%
\begin{equation*}
0=S_{n,\delta }-\left[ I_{n,\delta }+\Sigma _{\delta }+o_{p}\left( 1\right) %
\right] \sqrt{n}[(\hat{\delta}-\delta _{0})+\left[ I_{n,\delta }+\Sigma
_{\delta }+o_{p}\left( 1\right) \right] ^{-1}\mathbf{d]},
\end{equation*}%
with $\mathbf{d}=(0_{1\times d_{\gamma }},\{\left. \partial p\left(
b;\lambda \right) /\partial b\right\vert _{b=\beta
_{0,j}^{N}}\}_{j=1,...,d_{N}})^{\prime }$.\hfill $\square $

\begin{lemma}[Consistent model selection]
\label{lem 1}Assume that Assumptions 1-3 hold, and the condition (\ref{sign}%
) holds. If $\lambda \rightarrow 0$ and $\sqrt{n}\lambda \rightarrow \infty $
as $n\rightarrow \infty $, then, with probability tending to one for any $%
\delta =(\gamma ^{\prime },\beta ^{N\prime })^{\prime }$ satisfying $%
\left\Vert \delta -\delta _{0}\right\Vert =O_{p}(n^{-1/2})$, we have, for
any constant $c>0$, with $\theta ^{Z}=\left( \delta ^{\prime },0^{\prime
}\right) ^{\prime }$%
\begin{equation*}
Q_{n}\left( \theta ^{Z}\right) =\max_{\beta _{j}^{Z}\in \lbrack 0,\eta
_{j}n^{-1/2}]}Q_{n}\left( \theta \right) \text{,}
\end{equation*}%
for $0<\eta _{j}\leq c$ and $j=1,...,d_{Z}$.
\end{lemma}

\emph{Proof of Lemma \ref{lem 1}: }As in Fan and Li (2001, proof of Lemma
1), we establish, with $\delta -\delta _{0}=O_{p}\left( n^{-1/2}\right) $,
that the partial derivatives from the right satisfy, 
\begin{equation*}
\partial Q_{n}\left( \theta \right) /\partial \beta _{j}^{Z}<0\text{ for }%
0<\beta _{j}^{Z}<cn^{-1/2}\text{ with }j=d_{N}+1,...,d_{Z}\text{.}
\end{equation*}%
It holds that, 
\begin{align*}
\partial Q_{n}\left( \theta \right) /\partial \beta _{j}^{Z}& =\partial
L_{n}\left( \theta _{0}\right) /\partial \beta _{j}^{Z}+\left[ \partial
^{2}L_{n}\left( \theta _{0}\right) /\partial \beta _{j}^{Z}\partial \theta
^{\prime }\right] \left( \theta -\theta _{0}\right) \\
& +\left( \theta -\theta _{0}\right) ^{\prime }\partial ^{3}L_{n}\left(
\theta ^{\ast }\right) /\partial \beta _{j}^{Z}\partial \theta \partial
\theta ^{\prime }\left( \theta -\theta _{0}\right) /2-n\partial p\left(
\beta _{j}^{Z};\lambda \right) /\partial \beta _{j}^{Z},
\end{align*}%
with $\theta ^{\ast }$ between $\theta _{0}$ and $\theta $. Hence, by
Assumption \ref{Ass derivatives of lik}, and as $\theta =\left( \delta
^{\prime },(\beta ^{Z})^{\prime }\right) ^{\prime }$, with $\delta -\delta
_{0}=O_{p}\left( n^{-1/2}\right) $ and $0\leq \beta _{0}^{Z}\leq cn^{-1/2}$,
it follows that 
\begin{equation*}
\partial Q_{n}\left( \theta \right) /\partial \beta _{j}^{Z}=n\lambda _{n} 
\left[ O_{p}\left( n^{-1/2}/\lambda \right) -\lambda \partial p\left( \beta
_{j}^{Z};\lambda \right) /\partial \beta _{j}^{Z}\right] \text{.}
\end{equation*}%
Using (\ref{sign}) the result follows since by assumption $n^{-1/2}/\lambda
\rightarrow 0$. \hfill $\square $

\section{Golden Section Search \label{app: GSS}}

We here outline the Golden Section Search (GSS) algorithm from Kiefer (1953)
as applied to the modified information criterion $\IC^{\text{m}%
}\left( \lambda \right) $ in (\ref{eq: IC modified}).

The GSS algorithm is initialized by evaluating $\IC^{\text{m}%
}\left( \lambda \right) $ as a function of $\lambda \in \mathcal{A}_{0}$,
with 
\begin{equation*}
\mathcal{A}_{0}=\{a_{1}^{0},a_{2}^{0},a_{3}^{0},a_{4}^{0}\}=\{\lambda
_{0},(2-\varphi )\lambda _{m},(\varphi -1)\lambda _{m},\lambda _{m}\},
\end{equation*}%
where $\lambda _{0}=0$, $\lambda _{m}$ is the maximum value of $\lambda $
considered, chosen such that the $\hat{\beta}=0$, and the parameter $\varphi
=(1+\sqrt{5})/2$ is the so-called golden ratio.

If the value of $\IC^{\text{m}}\left( \lambda \right) $ is smallest
for $\lambda \in \{a_{1}^{0},a_{2}^{0}\}$, a new point $b^{0}=(2-\varphi
)a_{1}^{0}+(\varphi -1)a_{2}^{0}$ is introduced, and the set of points for $%
\lambda $ is updated to 
\begin{equation*}
\mathcal{A}_{1}=\{a_{1}^{1},a_{2}^{1},a_{3}^{1},a_{4}^{1}\}=%
\{a_{1}^{0},b^{0},a_{2}^{0},a_{3}^{0}\}.
\end{equation*}%
If, on the other hand, the $\IC^{\text{m}}\left( \lambda \right) $
is smallest for $\lambda \in \{a_{3}^{0},a_{4}^{0}\}$ a new point $%
c^{0}=(\varphi -1)a_{3}^{0}+(2-\varphi )a_{4}^{0}$ is introduced, and
instead $\mathcal{A}_{1}$ is defined by%
\begin{equation*}
\mathcal{A}_{1}=\{a_{1}^{1},a_{2}^{1},a_{3}^{1},a_{4}^{1}\}=%
\{a_{2}^{0},a_{3}^{0},c^{0},a_{4}^{0}\}.
\end{equation*}%
This iteration is repeated to obtain $\mathcal{A}_{2},\mathcal{A}_{3},...,%
\mathcal{A}_{M}$, where $M$ is classically set such that $\delta
_{M}=a_{4}^{M}-a_{1}^{M}$ is sufficiently small. As the $\IC^{\text{%
m}}\left( \lambda \right) $ depends on $\lambda $ only through the value of $%
\hat{d}_{\delta ,\lambda }$ (the estimated number of non-zero parameters
from the pQMLE)\textbf{, }the algorithm is here instead set to stop when $%
\hat{d}_{\delta ,a_{1}^{M}}-\hat{d}_{\delta ,a_{4}^{M}}\leq 1$. The final $%
\lambda $, $\lambda _{M}$ is then chosen as the point in $\mathcal{A}^{M}$
with the smallest $\IC^{\text{m}}\left( \lambda \right) $.

\newpage

\section*{References}

\begin{description}
\item \textsc{Ahrens, A., C.B. Hansen and M.E. Schaffer }(2020):
\textquotedblleft lassopack: Model Selection and Prediction with Regularized
Regression in Stata\textquotedblright , \emph{The Stata Journal}, 20(1),
176--235.

\item \textsc{Andrews, D.W.K. }(1999): \textquotedblleft Estimation when a
Parameter is on a Boundary\textquotedblright , \emph{Econometrica}, 67(6),
1341--1383.

\item ------ (2001): \textquotedblleft Testing when a Parameter is on the
Boundary of the Maintained Hypothesis\textquotedblright , \emph{Econometrica}%
, 69(3), 683--734.

\item \textsc{Belloni, A., and V. Chernozhukov} (2011): \textquotedblleft
High Dimensional Sparse Econometric Models: An
Introduction\textquotedblright . In: Alquier P., E. Gautier and G. Stoltz
(eds) \emph{Inverse Problems and High-Dimensional Estimation}. Lecture Notes
in Statistics, vol 203. Berlin: Springer.

\item \textsc{Bollerslev, T.} (1986):\ \textquotedblleft Generalized
Autoregressive Conditional Heteroskedasticity\textquotedblright , \emph{%
Journal of Econometrics}, 31, 307-327.

\item \textsc{Cavaliere, G., H.B. Nielsen} \textsc{and A.\ Rahbek }(2017):
\textquotedblleft On the Consistency of Bootstrap Testing for a Parameter on
the Boundary of the Parameter Space\textquotedblright , \emph{Journal of
Time Series Analysis, }38, 513--534.

\item \textsc{Cavaliere, G., H.B. Nielsen, R.S. Pedersen} \textsc{and A.\
Rahbek }(2022): \textquotedblleft Bootstrap Inference on the Boundary of the
Parameter Space, with Application to Conditional Volatility
Models\textquotedblright , \emph{Journal of Econometrics}, 227(1), 241--263.

\item \textsc{Cavaliere, G., I. Perera, } \textsc{and A.\ Rahbek }(2022):
\textquotedblleft Specification Tests for GARCH Processes with Nuisance
Parameters on the Boundary\textquotedblright , working paper, \emph{%
arXiv:2105.14081}.

\item \textsc{Fan, J. and R. Li }(2001): \textquotedblleft Variable
Selection via Nonconcave Penalized Likelihood and its Oracle
Properties\textquotedblright , \emph{Journal of the American Statistical
Association}, 96(456), 1348--1360.

\item \textsc{Francq, C. and J.M. Zako\"{\i}an }(2007): \textquotedblleft
Quasi-maximum Likelihood Estimation in GARCH Processes When Some
Coefficients are Equal to Zero\textquotedblright , \emph{Stochastic
Processes and their Applications}, 117(9), 1165--1372.

\item ------ (2009): \textquotedblleft Testing the Nullity of GARCH
Coefficients: Correction of the Standard Tests and Relative Efficiency
Comparisons\textquotedblright , \emph{Journal of American Statistical
Association}, 104, 313--324.

\item \textsc{Han, H. and D. Kristensen }(2014) \textquotedblleft Asymptotic
Theory for the QMLE in GARCH-X Models with Stationary and Nonstationary
Covariates\textquotedblright , \emph{Journal of Business and Economic
Statistics}, 32, 416-429.

\item \textsc{Kiefer, J.} (1953), \textquotedblleft Sequential Minimax
Search for a Maximum\textquotedblright , \emph{Proceedings of the American
Mathematical Society}, 4 (3): 502--506.

\item \textsc{Kopylev, L. and B. Sinha }(2010): \textquotedblleft On the
Asymptotic Distribution of Likelihood Ratio Test when Parameters Lie on the
Boundary\textquotedblright , Technical Report, Department of Mathematics and
Statistics, UMBC.

\item {\small ------\ }(2011): \textquotedblleft On the Asymptotic
Distribution of Likelihood Ratio Test when Parameters lie on the
Boundary\textquotedblright , \emph{Sankhy\={a} B,} 73, 20--41.

\item \textsc{Pedersen, R.S. and A. Rahbek }(2019): \textquotedblleft
Testing GARCH-X Type Models\textquotedblright , \emph{Econometric Theory},
35, 1012--1047.

\item \textsc{Tibshirani, R.} (1996), \textquotedblleft Regression Shrinkage
and Selection via the Lasso\textquotedblright , \emph{Journal of the Royal
Statistical Society}, 58, 267--288.
\end{description}

\end{document}